# Self Configurable Re-link Establishment using Continuous Neighbor Discovery in Asynchronous Sensor Networks

[1]Rushikesh B. Shreshtha, [2]Rajeswari Goudar

[1]Computer Department, MAEER'S MAE,
University of Pune 411015, Maharashtra, India

[2]Computer Department, MAEER'S MAE,
University of Pune 411015, Maharashtra, India

**Abstract**
A Sensor network generally has a large number of sensor nodes that are deployed at some audited site. In most sensor networks the nodes are static. Nevertheless, node connectivity is subject to changes because of disruptions in wireless communication, transmission power changes, or loss of synchronization between neighbouring nodes, so there is a need to maintain synchronization between the neighbouring nodes in order to have efficient communication. Hence even after a sensor is aware of its immediate neighbours, it must continuously maintain its view a process we call continuous neighbour discovery. In this proposed work we are maintaining synchronization between neighbouring nodes so that the sensor network will be always active.

## 1. Introduction

A sensor network may contain a huge number of simple sensor nodes that are deployed at some inspected site. In large areas, such a network usually has a mesh structure. In this case, some of the sensor nodes act as routers, forwarding messages from one of their neighbours to another. The nodes are configured to turn their communication hardware on and off to minimize energy consumption. Therefore, in order for two neighbouring sensors to communicate, both must be in active mode. In the sensor network model considered in this paper, the nodes are placed randomly over the area of interest and their first step is to detect their immediate neighbours the nodes with which they have a direct wireless communication and to establish routes to the gateway. In networks with continuously heavy traffic, the sensors need not invoke any special neighbour discovery protocol during normal operation. This is because any new node, or a node that has lost connectivity to its neighbours, can hear its neighbours simply by listening to the channel for a short time. However, for sensor networks with low and irregular traffic, a special neighbour discovery scheme should be used. Despite the static nature of the sensors in many sensor networks, connectivity is still subject to changes even after the network has been established. The sensors must continuously look for new neighbours in order to accommodate the following situations:

1) Loss of local synchronization due to accumulated clock drifts.
2) Disruption of wireless connectivity between adjacent nodes by a temporary event, such as a passing car or animal, a dust storm, rain or fog. When these events are over, the hidden nodes must be rediscovered.
3) The ongoing addition of new nodes, in some networks to compensate for nodes which have ceased to function because their energy has been exhausted.
4) The increase in transmission power of some nodes, in response to certain events, such as detection of emergent situations.

For these reasons, detecting new links and nodes in sensor networks must be considered as an ongoing process. We distinguished between detection of new links and nodes during initialization, i.e. when the node is in Init state, and their detection during normal operation. The former will be referred to as initial neighbour discovery whereas the latter will be referred to as continuous neighbour discovery. While previous works [1], [2], [3], [13], [15] address initial neighbour discovery and continuous neighbour discovery as similar tasks, to be performed by the same scheme, we claim that different schemes are required, for the following reasons: Initial neighbour discovery is usually performed when the sensor has no clue about the structure of its immediate surroundings. In such a case, the sensor cannot communicate with the gateway and is therefore very limited in performing its tasks. The immediate surroundings should be detected as soon as possible in order to establish a path to the gateway and contribute to the operation of the network. Hence in this state, more extensive energy use is justified [9],[12],[14]. In contrast, continuous neighbour discovery is performed when the sensor is already operational. This is a long term process, whose optimization is crucial for increasing network lifetime. When the sensor performs continuous neighbour discovery, it is already aware of most of its immediate neighbours and can therefore perform it together with these neighbours in order to consume less energy. In contrast, initial neighbour discovery must be executed by each sensor separately. Figure 1 shows a typical neighbour discovery protocol. In this protocol, a node becomes active according to its duty cycle. Let this duty cycle be in Init state and in Normal state. When a node becomes active, it transmits can invoke another procedure to finalize the setup of their joint wireless link. To summarize, in the Init state, a node has no information about its surroundings and therefore must remain active for a relatively long time in order to detect new neighbours. In contrast, in the normal state the node must use a more efficient scheme. Such a scheme is the subject of our study. When node 'u' is in the Init state, it performs initial neighbour discovery. After a certain time period, during which the node is expected, with high probability to most of its neighbours, the node moves to the Normal state, where continuous neighbour discovery is performed as shown in figure 2. A node in the Init state is also referred to in this paper as a hidden node and a node in the Normal state is referred to as a segment node.

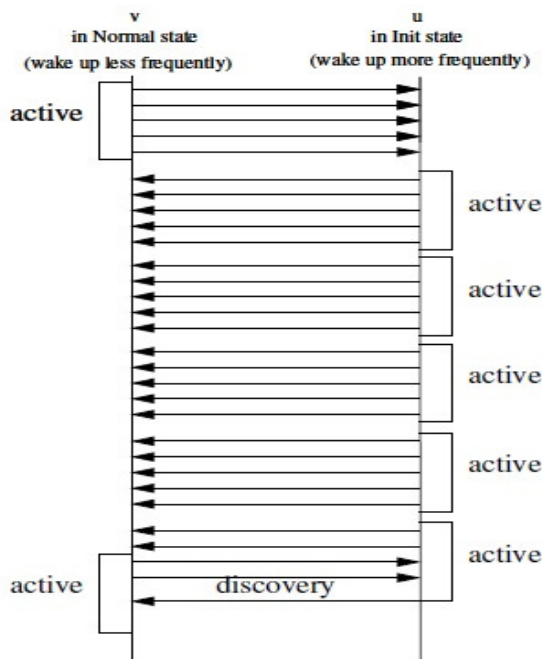

Figure 1. The transmission of HELLO messages in Init and Normal states

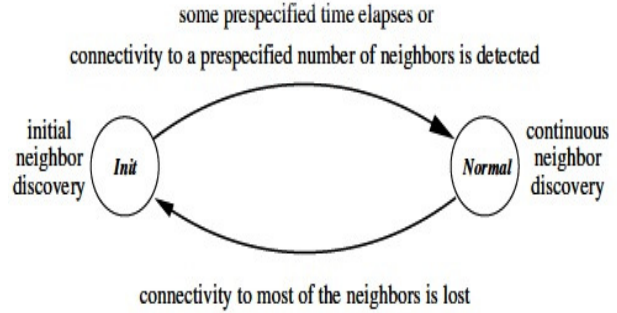

Figure 2. Continuous neighbour discovery vs. initial neighbour discovery in sensor networks

The main idea behind the continuous neighbour discovery scheme we propose is that the task of finding a new node 'u' is divided among all the nodes that can help node 'v' to detect node 'u'. These nodes are characterized as follows: (a) they are also neighbours of 'u' (b) they belong to a connected segment of nodes that have already detected each other; (c) node 'v' also belongs to this segment. Let degS (u) be the number of these nodes. This variable indicates the in-segment degree of a hidden neighbour 'u'. In order to take advantage of the proposed discovery scheme, node 'v' must estimate the value of degS (u).

## 2. Related Work

In a special node, called an access point, we are using this point in Wi-Fi network operating in centralized node. The Messages are transmitted only to or from the point. In the process of neighbour discovery, a new node can be detected by the base station. Discovering the new node is easy when compared the energy consumption is not a concern for the base station. The base station broadcasts a special HELLO message1. This message can hear that particular regular node to initiate a registration process. The regular node can switch frequencies/channels in order to handle the best HELLO message for its needs. This is the best message that might be depending on the identity of the broadcasting base station, on security considerations. All these problems related the collisions of messages in such a network are addressed in [4], [10], [11]. So other works trying to minimize the discovery time by optimizing the broadcast rate of the HELLO messages [1], [5], [6], [7], [8].

## 3. Basic Scheme and Problem

We assume that all nodes are having the same transmission range, it means for every time the connectivity is always bi-directional. In our analysis, the network is a unit disk graph; means: the pair of the nodes that can be within

47

transmission range are should be neighbouring nodes. These two nodes are said to be directly connected, and are aware of each other's wake-up times. Two nodes are said to be connected if there is a path of directly connected nodes between them. A group of connected nodes is known as a segment. Consider a pair of neighbouring nodes that belong to the same segment but are not aware that they have direct wireless connectivity.

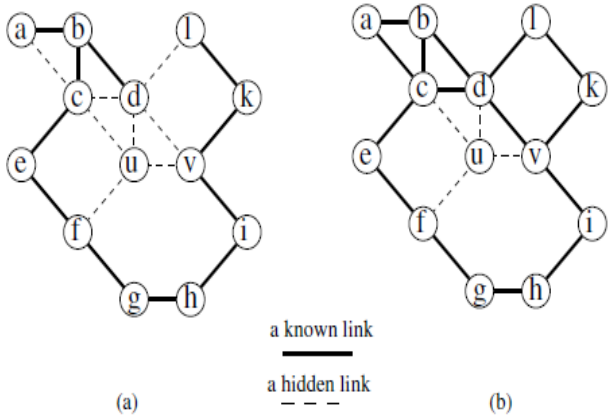

Figure 3. Segments with hidden nodes and links

In figure 3 the node 'c' can learn about their hidden wireless link using the following simple scheme, which uses two message types:

(a) SYNC messages for synchronization between all segment nodes, transmitted over known wireless links.

(b) HELLO messages for detecting new neighbours.

*Scheme 1 (detecting all hidden links inside a segment):*

Whenever a new node is discovered by one of the segment nodes it can detect all hidden links inside a segment. For all segment members, the discovering node issues a special SYNC message asking them to periodically broadcast a group of HELLO messages. The SYNC message is passes over the already known wireless links of the segment. So every segment node has to be
guaranteed to be received.

*Scheme 2 (detecting a hidden link outside a segment):*

In this scheme, the same segment is used to minimize the possibility of repeating collisions between the HELLO messages of nodes. Practically, another scheme might be used, where segment nodes coordinate their wake-up periods for prevents collisions. However, finding an efficient time division is equivalent to the well-known node colouring problem, which is node 'u' wakes up randomly.

The value of T(u) is as follows:

T(u) = TI , if node u is in the Init state
T(u) = TN(u), if node u is in Normal state

## 4. Proposed Method

As already explained, we consider the discovery of hidden neighbours as a joint task to be performed by all segment nodes. We need to estimate the number of in-segment neighbours of every hidden node u, denoted by degS(u) to determine the discovery load to be imposed on every segment node namely how often such a node should become active and send HELLO messages, In this section, 'I' presents methods that can be used by node 'v' in the Normal state to calculate this value. Node 'u' is assumed to not yet be connected to the segment and it is in the Init (initial neighbour discovery) state. Here first we have to measures node 'v', the average in-segment degree of the segment's nodes, we have to use this number as an estimate of the in-segment degree of 'u'. The average in-segment degree of the segment's nodes can be calculated by the segment leader. The end of this, it gets from every node in the segment and immediately a message indicating the in-segment degree of the sending node, which is known due to Scheme node 'v' discovers, using Scheme 1, the number of its in-segment neighbours, degS(v), and views this number as an estimate of degS(u). When the degrees of neighbouring nodes are strongly correlated, this approach will give good results than the previous one. Node 'v' uses the average in-segment degree of its segment's nodes and its own in-segment degree degS(v). To estimate the number of node u's neighbours. This approach gives the best results if the correlation between the in-segment degrees of neighbouring nodes is known.

## 5. Conclusion

We exposed a new problem in wireless sensor networks, referred to as ongoing continuous neighbor discovery. We argue that continuous neighbor discovery is crucial even if the sensor nodes are static. If the nodes in a connected segment work together on this task, hidden nodes are guaranteed to be detected within a certain probability P and a certain time period T, with reduced expended on the detection. We proposed that our scheme works well if every node connected to a segment estimates the in-segment degree of its possible hidden neighbors and continuous neighbor discovery algorithm determines the

48

frequency with which every node enters the HELLO period.